\documentclass[nofootinbib,prd,preprintnumbers,superscriptaddress,aps]{revtex4}
\usepackage{amsmath}
\usepackage{amssymb}
\usepackage{amsfonts}
\usepackage{subcaption}
\usepackage{graphicx}
\usepackage[colorlinks=true,linkcolor=blue,citecolor=teal,urlcolor=blue]{hyperref}
\usepackage{xcolor}
\usepackage{hyperref}

\usepackage{physics}
\usepackage{bbold}
\usepackage{float}

\usepackage{cases}
\usepackage{braket}
\usepackage[T1]{fontenc}
\usepackage{mathrsfs}
\usepackage{enumerate}
\usepackage{bm}
\usepackage{multirow}
\usepackage{comment}
\usepackage{mathrsfs}

\makeatother
 
\begin{document}
 \title{From information bounds to infrared gravity: implications of Sharma--Mittal entropy}

	\author{Abdelhakim Benkrane}   
	\email{abdelhakim.benkrane@univ-ouargla.dz, hakim9502.benkrane@gmail.com}	
\affiliation{Université Kasdi Merbah Ouargla, Laboratoire LRPPS, Ouargla 30000, Algeria}

\author{Giuseppe Gaetano Luciano}
\email{giuseppegaetano.luciano@udl.cat}
\affiliation{Department of Chemistry, Physics and Environmental and Soil Sciences, Polytechnic School, University of Lleida, Av. Jaume II, 69, 25001 Lleida, Spain
}

\author{Ahmad Sheykhi}
\email{asheykhi@shirazu.ac.ir}
\affiliation{Department of Physics, College of Science, Shiraz University, Shiraz 71454, Iran}
\affiliation{Biruni Observatory, College of Science, Shiraz University, Shiraz 71454, Iran}

%%%%%%%%%%%%%%%%%%%%%%%%%%%%%%%%%%%%%%%%%%%%%%%%%%%%%%%%%%%%%%%%%%%%%%%%%%%%%%%%%%%%%%%%%%%%%%%%

\begin{abstract}
The Sharma--Mittal (SM) entropy provides a two-parameter generalization encompassing both the R\'{e}nyi and Tsallis statistical frameworks. In this work, we investigate its thermodynamic and gravitational implications in the context of black hole physics and emergent gravity. Specifically, we examine the compatibility of the gravitational realization of the SM entropy with the Bekenstein bound and show that the corresponding framework consistently interpolates between the R\'{e}nyi and Bekenstein--Hawking entropies in the appropriate limits. By incorporating Landauer's principle into black hole thermodynamics, we obtain a modified mass-loss relation associated with one-bit information erasure, exhibiting nontrivial parameter-dependent asymptotic behavior in both the small- and large-mass regimes. Furthermore, within Verlinde's entropic gravity framework, we derive modified gravitational force and acceleration laws induced by the SM entropy. We show that the resulting acceleration deviates from the Newtonian prediction at large distances and naturally reproduces a MOND-like regime for the specific parameter relation $R/\delta = 3/2$. This condition establishes a direct connection between the SM entropy parameters and the MOND acceleration scale $a_0$. Our findings highlight the potential of the SM framework to provide a unified link between black hole thermodynamics, information theory, and infrared modifications of gravity, while offering new insights into phenomena traditionally attributed to the dark matter paradigm.
\end{abstract}

\date{\today}
\maketitle
%%%%%%%%%%%%%%%%%%%%%%%%%%%%%%%%%%%%%%%%%%%%%%%%%%%%%%%%%%%%%%%%%%%%%%%%%%%%%%%%%%%%%%%%%%%%%%%%%%%%%%%%%%%%%%%%
\section{Introduction}

The discovery of black hole thermodynamics revealed a profound  connection between gravitation, thermodynamics and quantum theory. The identification of black hole entropy and temperature~\cite{Bekenstein1973,Hawking1975}, together with the derivation of the Einstein field equations from the Clausius relation~\cite{Jacobson:1995ab}, suggested that gravitational dynamics may possess an intrinsically thermodynamic origin. This perspective has been further developed in a variety of contexts, including the thermodynamics of cosmological horizons~\cite{Bak2000,CaiKim2005,AkbarCai2007,Frolov:2002va,Padmanabhan:2009vy}, the interpretation of gravity as an emergent phenomenon~\cite{Padmanabhan:2009vy,verlinde2011origin}, and the investigation of the microscopic degrees of freedom underlying spacetime.

Despite these remarkable developments, the thermodynamic description of gravitating systems remains a subject of ongoing debate. Motivated by the still unknown microphysical structure of spacetime, the intrinsically long-range nature of gravitational interactions, and the non-extensive character of the Bekenstein--Hawking entropy~\cite{Touchette2002,Biro2011,Biro2013,Tsallis2013,Majhi2017}, several generalized entropy formalisms have been investigated as an alternative route to modified gravity scenarios \cite{CapEx}, aiming to account for gravitational phenomena through minimal modifications of the standard thermodynamic framework rather than altering the underlying gravitational dynamics. Among these, the R\'{e}nyi and Tsallis entropies~\cite{Touchette2002,Biro2011,Majhi2017,Tsallis2011,Renyi:1961,Tsallis1988,Abe2001,GellMannTsallis2004,AbeOkamoto2001,Abe2006,Taruya2003} have attracted considerable attention and have been employed to investigate a wide range of problems in gravitation and cosmology~\cite{Biro2013,Tsallis2013,Majhi2017,Taruya2003,Leubner2004,Plastino1993,Lima2002,Hansen2005,Leubner2005,Dey2016,Bialas2008,Luciano:2025hjn,Czinner2016,Guo2014,Komatsu2017,Moradpour2017,Moradpour2016c,Luciano:2025elo,Abreu2013,Saridakis:2018unr,Capozziello:2025axh,Lymperis:2018iuz,Luciano:2026vhm,GINE2026140492}.

Another notable extension is provided by the Sharma--Mittal (SM) entropy~\cite{Sharma1975,Sharma1977}, characterized by two independent deformation parameters. By encompassing both the R\'{e}nyi and Tsallis entropies as limiting cases, the SM entropy offers a flexible framework to investigate possible departures from the standard entropy-area relation and their implications for gravitational phenomena~\cite{jahromi2018generalized}.

Within this thermodynamic scenario, the role of information in gravitational systems has assumed increasing significance. Since horizon entropy is often interpreted as encoding information associated with degrees of freedom inaccessible to a given observer, it is natural to ask whether the physical principles governing information processing may also have gravitational consequences. In this context, the Landauer principle provides a compelling bridge between information theory and horizon thermodynamics. According to this principle, the irreversible erasure of one bit of information requires a minimum energy dissipation of $k_{\mathrm{B}}T\ln 2$ into the environment~\cite{landauer1961irreversibility,Bennett1982,Parrondo2015}. By assigning an energetic content to information, the Landauer principle establishes a fundamental thermodynamic limit on irreversible information processing and emphasizes that information cannot be regarded as a purely abstract entity~\cite{Parrondo2015}.

In gravitational settings, the energetic implications of the Landauer principle naturally depend on the entropy functional adopted to characterize the underlying microscopic degrees of freedom.
From this perspective, generalized entropy formalisms are not merely alternative statistical descriptions, but may lead to distinct thermodynamic contributions arising from information processing. This observation provides an additional motivation for investigating the consequences of the SM entropy in black hole thermodynamics and emergent gravitational dynamics~\cite{abreu2025modified}.

Starting from these premises, in this work we investigate several thermodynamic and gravitational implications of the SM entropy formalism. We first review the generalized entropic framework and the main properties of the SM entropy in Sec.~\ref{GEF}. In Sec.~\ref{Bb}, we investigate the compatibility of the gravitational realization of the SM entropy with the Bekenstein bound and discuss the corresponding limiting behaviors. We then explore, in Sec.~\ref{LP}, the consequences of incorporating the Landauer principle into black hole thermodynamics, deriving the modified mass-loss relation associated with one-bit information erasure. Finally, in Sec.~\ref{EMG}, we investigate the implications of the SM entropy within Verlinde's entropic gravity framework by deriving modified gravitational force and acceleration laws and analyzing their asymptotic behavior. In particular, we show that the resulting acceleration naturally gives rise to a MOND-like regime under suitable constraints on the SM parameters. These findings highlight the potential of generalized entropy formalisms to bridge black hole thermodynamics, information theory, and emergent gravity. Unless otherwise stated, natural units are assumed throughout this work.

%%%%%%%%%%%%%%%%%%%%%%%%%%%%%%%%%%%%%%%%%%%%%%%%%%%%%%%%%%%%%%%%%%%%%%%%%%%%%%%%%%%%%%%
\section{Generalized Entropy Formalism}
\label{GEF}
The generalized entropy frameworks proposed by R\'{e}nyi and
Tsallis represent two important one-parameter extensions of the
standard Boltzmann-Gibbs entropy
\cite{Tsallis2011,Renyi:1961,Tsallis1988}. They are respectively
expressed as
\begin{equation}
S_{\mathrm{R}}=\frac{1}{\delta}\ln\left(\sum_{i=1}^{W}P_i^{\,1-\delta}\right),
\qquad
S_{\mathrm{T}}=\frac{1}{\delta}\left(\sum_{i=1}^{W}P_i^{\,1-\delta}-1\right),
\label{tsallis}
\end{equation}
where $P_i$ denotes the probability associated with the $i$-th microstate among the total number of $W$ accessible states, subject to the normalization condition $\sum_{i=1}^{W} P_i = 1$. 
In addition, $\delta$ is a real parameter quantifying the deviation from the standard entropic framework. In the limit $\delta\rightarrow 0$, both the R\'{e}nyi and Tsallis entropies consistently recover the standard Boltzmann--Gibbs--Shannon entropy. Furthermore, for a given value of the deformation parameter, these two entropy measures are related through the exact relation $S_{\mathrm{R}}=
\frac{1}{\delta}
\ln\left(1+\delta S_{\mathrm{T}}\right)$~\cite{Tsallis2011,Renyi:1961,Tsallis1988,Komatsu2017,Moradpour2017}, which shows that the R\'{e}nyi entropy can be expressed as a monotonic function of the Tsallis entropy.

A broader entropy measure was introduced by Sharma and Mittal as a two-parameter generalization encompassing both the R\'{e}nyi and Tsallis entropies as limiting cases~\cite{Masi2005,Sharma1975,Sharma1977,Frank2000,FrankPlastino2002}. The Sharma--Mittal (SM) entropy is defined as
\begin{equation}
    S_{\mathrm{SM}}
    =
    \frac{1}{1-r}
    \left[
    \left(
    \sum_{i=1}^{W}P_i^{\,1-\delta}
    \right)^{\frac{1-r}{\delta}}
    -1
    \right],
    \label{SM}
\end{equation}
where $r$ denotes an additional free real parameter characterizing the generalized structure of the entropy. The SM entropy consistently recovers the R\'{e}nyi entropy in the limit $r\to 1$, while the Tsallis entropy is obtained for $r=1-\delta$.

Several studies have investigated the mathematical and physical properties of the SM entropy and demonstrated its applicability to a wide class of statistical systems~\cite{Masi2005,Sharma1975,Sharma1977,Frank2000,FrankPlastino2002}. Owing to its ability to interpolate between different non-extensive scenarios through two independent deformation parameters, Eq. \eqref{SM} provides a particularly suitable framework for probing the consequences of generalized statistical descriptions in situations where the underlying microscopic degrees of freedom remain unknown, as is expected in gravitational systems.

Using Eq.~(\ref{tsallis}), the SM entropy can be conveniently rewritten as
\begin{equation}
S_{\mathrm{SM}}
=
\frac{1}{R}
\left[
\left(1+\delta S_{\mathrm T}\right)^{R/\delta}
-1
\right],
\label{pro}
\end{equation}
where, for convenience, we have introduced the parameter
\(
R\equiv 1-r
\).

Recent studies have suggested that the non-additive character of the Bekenstein-Hawking entropy may be naturally interpreted within the Tsallis statistical framework (see Ref.~\cite{jahromi2018generalized} and the references therein). Within this picture, the Bekenstein--Hawking entropy is regarded as the Tsallis entropy associated with the underlying gravitational degrees of freedom. Consequently, the SM entropy acquires a direct gravitational realization, namely~\cite{jahromi2018generalized}
\begin{gather}
S_{\mathrm{SM}}=\dfrac{1}{R}
\left[
\left(1+\delta S_{\mathrm{BH}}\right)^{R/\delta}
-1
\right].
\label{pro1}
\end{gather}
In the following sections, we investigate the thermodynamic and gravitational consequences of this generalized entropy, focusing on the Bekenstein bound, the Landauer principle, and emergent gravity.

%%%%%%%%%%%%%%%%%%%%%%%%%%%%%%%%%%%%%%%%%%%%%%%%%%%%%%%%%%%%%%%%%%%%%%%%%%%%%%%%%%%%%%%%%%%%%%%%%%%%%%%%%%%%%%%%%%%%%%%%%%%%%%%%%%%%%%%
\section{Bekenstein bound conjecture}
\label{Bb}

Black hole thermodynamics has brought to light an intriguing relationship between entropy, spacetime geometry and information, suggesting that spacetime itself may encode fundamental information-theoretic content. Within this perspective, it is natural to investigate the fundamental limits governing the entropy of physical systems.

In this context, Bekenstein derived a general upper bound on the entropy of a finite quantum system~\cite{bekenstein1981universal},
\begin{equation}
S \leq \frac{2\pi k_B R E}{\hbar c},
\end{equation}
where we have temporarily restored the fundamental constants for clarity. This relation constrains the maximal entropy of a system in terms of its energy $E$ and characteristic size $R$. In particular, for a Schwarzschild black hole one has $E=M$ and $R=2M$, yielding $S \leq \pi R^2 = S_{\mathrm{BH}}$.

Assuming that the gravitational realization of the SM entropy should also satisfy the Bekenstein bound, $S_{\mathrm{SM}} \leq S_{\mathrm{BH}}$,
and using Eq.~(\ref{pro1}), we obtain
\begin{equation}
1 \leq
\dfrac{\left(RS_{\mathrm{SM}}+1\right)^{\delta/R}-1}
{\delta\, S_{\mathrm{SM}}}\,,
\label{Beknstein}
\end{equation}
which represents the consistency condition required for compatibility with the Bekenstein bound.

Equation \eqref{Beknstein} establishes a nontrivial constraint controlled by the deformation parameters $R$ and $\delta$, thereby identifying the region of parameter space in which the SM framework remains consistent with the Bekenstein bound. The corresponding limiting cases consistently recover the previously known results, reinforcing the internal consistency of the formalism. The existence of such constraints suggests that generalized entropy corrections cannot be chosen arbitrarily, but must satisfy fundamental thermodynamic requirements inherited from black hole physics.

To determine the conditions under which the compatibility condition is fulfilled, we introduce the  variables \begin{equation} 
x = R S_{\mathrm{SM}}, \qquad y = \delta S_{\mathrm{SM}}\,. 
\end{equation} 
The admissible parameter space is  analyzed by numerically determining the boundary curve $y(x)$ associated with Eq.~(\ref{Beknstein}). The region satisfying the compatibility condition is identified by shading the corresponding allowed domain in the $(x,y)$ plane.

\begin{figure}[t]
    \centering
    \begin{subfigure}[t]{0.45\linewidth}
        \centering
        \includegraphics[width=\linewidth,height=0.25\textheight]{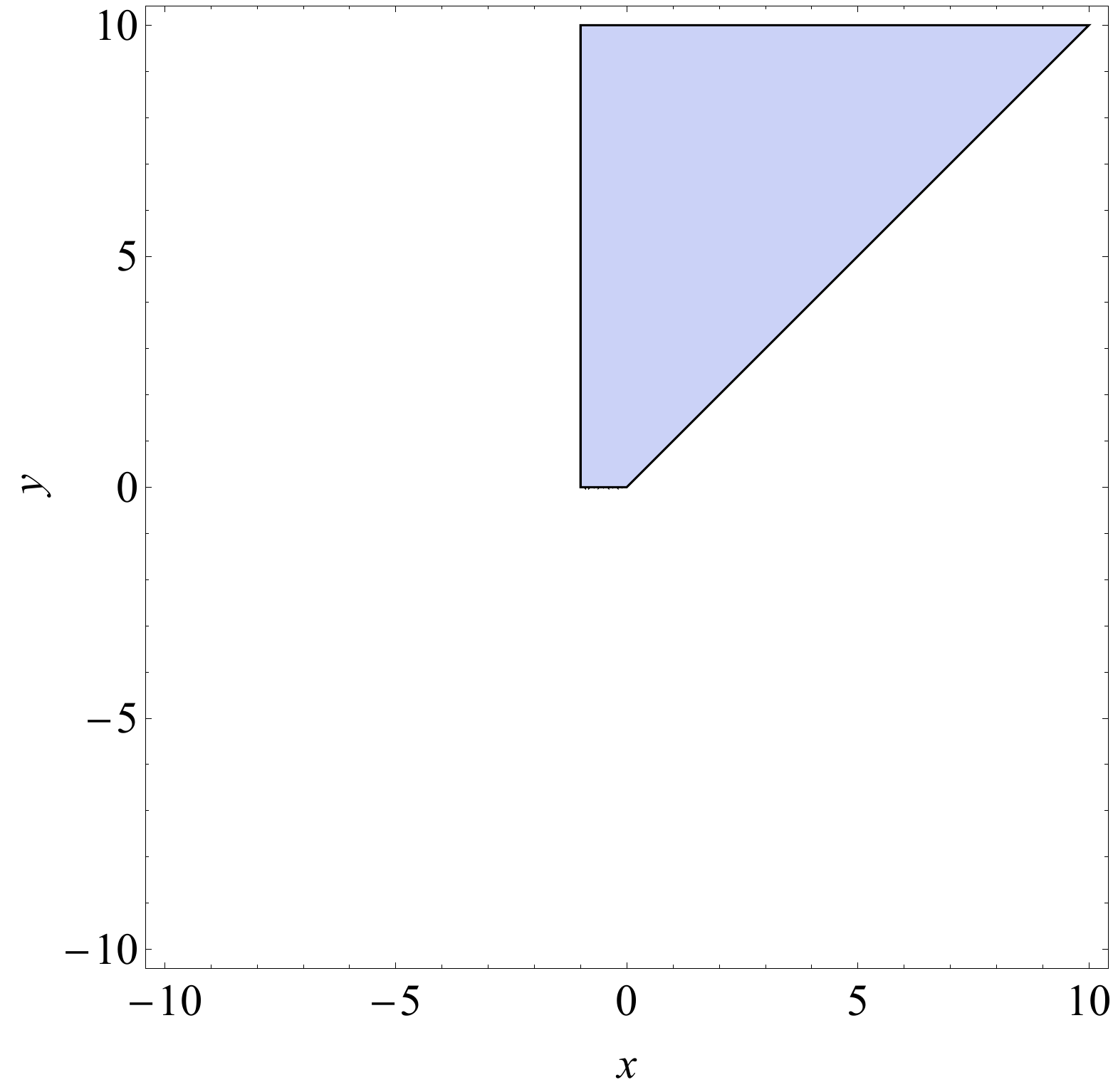}
        \caption{$y>x$.}
        \label{BCONDITIONa}
    \end{subfigure}
    \hfill
    \begin{subfigure}[t]{0.45\linewidth}
        \centering
        \includegraphics[width=\linewidth,height=0.25\textheight]{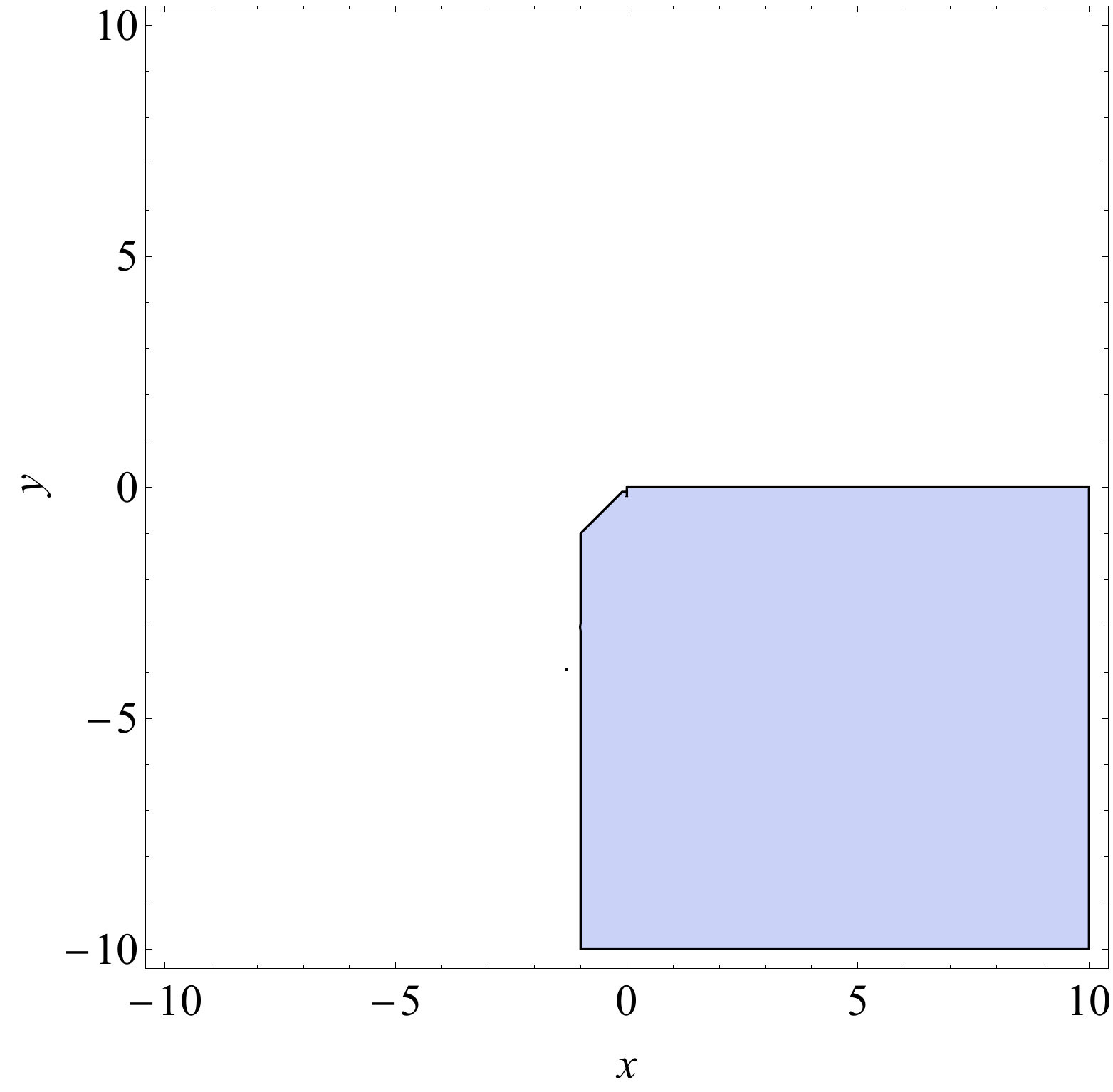}
        \caption{$x>y$.}
        \label{BCONDITIONb}
    \end{subfigure}
    \caption{
    Admissible regions associated with the compatibility condition~(\ref{Beknstein}) in the SM framework. Panel (a) shows the allowed domain for $y>x$, while panel (b) corresponds to the additional constraint $x>y$. The shaded regions identify the parameter combinations for which the SM entropy remains compatible with the Bekenstein bound.
    }
    \label{BCONDITION}
\end{figure}

In Fig.~\ref{BCONDITION}, we illustrate the admissible regions associated with the condition~(\ref{Beknstein}) in the $(x,y)$ parameter space. Specifically, the left panel displays the allowed domain in the case $y>x$. The admissible region is confined to the first quadrant and is bounded by a positively sloped curve. As $x$ increases, the minimum allowed value of $y$ also increases, indicating that the deformation parameters are subject to correlated constraints if compatibility with the Bekenstein bound is to be maintained. Only a restricted subset of the parameter space satisfies the required condition.

The right panel, on the other hand, shows the admissible region under the additional constraint $x>y$. In this case, within the plotted range, the allowed domain is predominantly characterized by positive values of $x$ and non-positive values of $y$. This behavior illustrates how additional restrictions on the deformation parameters further narrow the set of viable SM configurations compatible with the Bekenstein bound.

It is worth noting that the regions corresponding to $y<0$, and hence to negative values of the deformation parameter $\delta$, are mathematically admissible within the present compatibility analysis. This does not imply, however, that such configurations remain physically viable in all contexts. Indeed, once the explicit black hole realization of the SM entropy is invoked, additional requirements may arise from the condition that the generalized entropy remain real. As will be discussed in the next Section, this leads to further restrictions on the admissible parameter space when analyzing the black hole mass-loss relation.

Overall, the two panels demonstrate that the SM deformation parameters are not arbitrary, but are subject to nontrivial constraints imposed by the requirement of compatibility with the Bekenstein bound. This provides an additional consistency test for the gravitational realization of the generalized entropy formalism.

In the limit $R\rightarrow 0$, the  condition \eqref{Beknstein} reproduces the corresponding result obtained within the R\'{e}nyi entropy framework~\cite{abreu2025modified}, namely,
\begin{gather}
1\leq
\dfrac{\exp(\delta S_{\mathrm{SM}})-1}
{\delta S_{\mathrm{SM}}}.
\end{gather}
This limiting case confirms the consistency of the present approach with the R\'{e}nyi formulation, showing that the SM framework naturally recovers the previously established compatibility condition. As discussed in Ref.~\cite{abreu2025modified}, in the limit $\delta\to0$ the corresponding ratio approaches unity, whereas for increasing values of $\delta$ it remains larger than one and increases monotonically, thereby preserving compatibility with the Bekenstein bound.

On the other hand, in the limit $R\rightarrow\delta$, the condition reduces to the trivial identity $1\leq1$.
This reflects the recovery of the standard Bekenstein--Hawking entropy within the gravitational realization of the SM framework. In this regime, the compatibility condition is identically satisfied, indicating that no additional restrictions arise beyond those already encoded in the conventional semiclassical description of black hole entropy.

We would like to mention that, although our analysis has been carried out within the standard Bekenstein bound, generalized versions incorporating quantum-gravity-induced modifications of the uncertainty principle have also been considered~\cite{Medved:2004yu,Buoninfante:2020guu,Ong:2025ent}. Investigating the robustness of the present compatibility condition in such scenarios constitutes an interesting direction for future work.

\section{Landauer principle}
\label{LP}

As discussed in the Introduction, the Landauer principle establishes a fundamental thermodynamic cost associated with irreversible information processing. In particular, the erasure of one bit of information necessarily requires a minimum energy dissipation given by $    \Delta E \geq T \ln 2$, 
where $T$ is the temperature of the system~\cite{landauer1961irreversibility}. The factor $\ln 2$ originates from the binary nature of the erased information and reflects the entropy change associated with reducing the number of accessible microstates by a factor of two.

In the context of black hole thermodynamics, this relation provides a natural framework to investigate the energetic consequences of information erasure when generalized entropy functionals are employed. In what follows, we examine how the SM entropy modifies the corresponding mass-loss relation induced by the deletion of a single bit of information.

For a Schwarzschild black hole, the Bekenstein--Hawking entropy is $S_{\mathrm{BH}}=4\pi M^{2}$. We  consider an elementary process in which the black hole undergoes an infinitesimal mass variation associated with the loss of one bit of information. Differentiating $S_{\mathrm{BH}}$ gives $\Delta S_{\mathrm{BH}}=8\pi M\Delta M$, 
where higher-order terms in $\Delta M$ have been neglected.

To investigate the impact of generalized entropy on the information-erasure process, we consider the SM extension of the Bekenstein--Hawking entropy introduced in Eq.~(\ref{pro1}). The corresponding change in entropy can be written as
\begin{equation}
\Delta S_{\mathrm{SM}}
=
\Delta S_{\mathrm{BH}}
\left(1+\delta S_{\mathrm{BH}}\right)^{R/\delta-1}.
\label{dSSM}
\end{equation}
Invoking the Landauer principle, the erasure of one bit of information is associated with the entropy variation
$\Delta S_{\mathrm{SM}}=\ln 2$. 
Combining this relation with Eq.~(\ref{dSSM}), we obtain the corresponding black hole mass variation,
\begin{equation}
\Delta M=
\dfrac{\ln 2}{8\pi M}
\left(1+4\pi\delta M^2\right)^{1-R/\delta}.
\label{var}
\end{equation}
This expression establishes a direct connection between information erasure and black hole mass variation within the SM framework. The modified mass-loss relation shows that the thermodynamic response of black holes to one-bit information erasure depends sensitively on the underlying entropy functional, with deviations from the standard semiclassical result controlled by the deformation parameters $R$ and $\delta$.

As expected, the relevant limiting cases are consistently recovered: the standard Bekenstein--Hawking result, $\Delta M=\frac{\ln 2}{8\pi M}$,
is obtained for $R=\delta$, whereas the corresponding R\'enyi expression is reproduced in the limit $R\to0$~\cite{abreu2025modified}. This provides an additional consistency check of the generalized formalism.

It is worth noting that, although both positive and negative values of the deformation parameter $\delta$ are formally admissible within the generalized entropy framework, in the following we restrict our analysis to the case $\delta>0$. Indeed, for $\delta<0$, the factor $(1+4\pi\delta M^2)$ appearing in Eq.~(\ref{var}) remains positive only for $M<\frac{1}{\sqrt{4\pi|\delta|}}$,
thereby restricting the range of admissible black hole masses. By contrast, the choice $\delta>0$ ensures that the generalized mass-loss relation is well defined for arbitrary black hole masses and allows for a straightforward interpretation of the resulting evaporation regimes.

Under the assumption $\delta>0$, Eq.~(\ref{var}) allows one to identify qualitatively distinct evaporation behaviors. In particular, the sign of the exponent $1-R/\delta$ determines whether the generalized entropy suppresses or enhances the black hole mass variation relative to the Bekenstein--Hawking case. Specifically, the mass variation is suppressed for $R/\delta>1$, whereas it is enhanced for $R/\delta<1$. Therefore, SM corrections may either inhibit or amplify the evaporation process depending on the parameter regime.

To gain further insight into the physical implications of the generalized framework, let us consider the small-mass regime, characterized by $4\pi\delta M^2\ll1$. Expanding Eq.~(\ref{var}) to first nontrivial order, one finds
\begin{equation}
\Delta M
\simeq
\frac{\ln 2}{8\pi M}
\left[
1+4\pi(\delta-R)M^2
\right].
\end{equation}
This result shows that the first correction to the standard semiclassical expression is governed by the combination $(\delta-R)$. Therefore, the leading departure from the Bekenstein--Hawking prediction depends only on the mismatch between the two deformation parameters rather than on their individual values. In the particular case $\delta=R$, the standard behavior is recovered.

Let us now turn to the opposite regime of sufficiently large black hole masses, $4\pi\delta M^2\gg1$.
In this limit, the unity term inside the bracket of Eq.~(\ref{var}) can be neglected, and the mass variation becomes
\begin{gather}
\Delta M
\simeq
\frac{\ln2}{8\pi}
\left(4\pi\delta\right)^{1-R/\delta}
M^{\,1-2R/\delta}.
\end{gather}
The asymptotic behavior is therefore governed by the ratio $R/\delta$. For $R/\delta<1/2$, the mass variation increases with the black hole mass, whereas for $R/\delta>1/2$ it decreases. The critical value $R/\delta=1/2$ separates these two regimes and yields a mass-independent asymptotic value of $\Delta M$. This critical behavior indicates that the energetic cost associated with one-bit information erasure is not universal in the large-mass regime, but depends sensitively on the underlying generalized entropy through the ratio $R/\delta$.

%%%%%%%%%%%%%%%%%%%%%%%%%%%%%%%%%%%%%%%%%%%%%%%%%%%%%%%%%%%%%%%%%%%%%%%%%%%%%%%%%%%%%%%%%%%%%%%%%%%%%%%%%%%%%%%%%%%%%%%%%%%%%%%%%%%%%%%%%%%%%%%%%%%%

\section{Emergent Gravity from Sharma-Mittal Entropy}
\label{EMG}
The possibility that gravity may emerge from thermodynamic and informational principles has attracted considerable attention in recent years. In Verlinde's proposal~\cite{verlinde2011origin}, gravity is interpreted as an emergent phenomenon arising from entropy variations on holographic screens rather than as a fundamental interaction. In this framework, modifications of the underlying entropy--area relation naturally induce corrections to the corresponding gravitational dynamics.

In the present context, the gravitational realization of the SM entropy provides a natural setting in which to investigate how generalized statistical effects modify the gravitational force law. To this end, within the generalized entropic gravity framework, modifications of the entropy--area relation can be incorporated through the expression for the magnitude of the entropic force~(see, e.g., Refs.~\cite{Sheykhi:2010wm,benkrane2025entropy}),
\begin{equation}
F=
\frac{GMm}{r^{2}}\,
4\ell_{p}^{2}
\frac{dS}{dA},
\label{entropicforce}
\end{equation}
where $A$ denotes the area of the holographic screen and $S$ is the associated entropy. The quantity $dS/dA$ therefore encodes the deviations from the standard Newtonian force induced by the underlying generalized entropy. In contrast to the previous sections, the fundamental constants are restored here in order to make the role of the Planck scale explicit. The fundamental quantum-gravitational scale entering this construction is the Planck area, $\ell_p^2=G\hbar/c^3$, which sets the natural unit for holographic entropy relations. For the standard Bekenstein--Hawking entropy, $S=A/(4\ell_p^2)$, Eq.~(\ref{entropicforce}) reduces to the Newtonian law of gravity.

We now specialize the entropic-force framework to the gravitational realization of the SM entropy introduced in Eq.~(\ref{pro1}). Substituting the corresponding entropy--area relation into Eq.~(\ref{entropicforce}), we obtain the modified entropic force
\begin{equation}
    F_{\text{SM}}
    =
    \frac{GMm}{r^{2}}
    \left(
    1+\delta S_{\text{BH}}
    \right)^{R/\delta-1}=\frac{GMm}{r^{2}}
    \left(
    1+
    \delta\frac{\pi r^{2}}{\ell_p^{2}}
    \right)^{R/\delta-1}\,,
    \label{FSMfinal}
\end{equation}
where, in the second equality, we have used the explicit expression of the Bekenstein--Hawking entropy in terms of the holographic screen radius. 

As expected, the known limiting cases are consistently recovered. For $R=\delta$, Eq.~(\ref{FSMfinal}) reduces to the standard Newtonian force law, while in the limit $R\to0$ it reproduces the corresponding R\'enyi-type correction~\cite{abreu2025modified},
\begin{equation}
F_{\mathrm{R}}=
\frac{GMm}{r^{2}}
\frac{1}{1+\delta \frac{\pi r^{2}}{\ell_{p}^{2}}}.
\end{equation}

The corresponding acceleration is obtained by dividing the modified force, Eq.~(\ref{FSMfinal}), by the test mass. To quantify the deviations from the Newtonian prediction, it is convenient to consider the ratio
\begin{equation}
\frac{a_{\mathrm{SM}}}{a_{\mathrm{N}}}
=
\left(
1+
\delta\frac{\pi r^{2}}{\ell_p^{2}}
\right)^{R/\delta-1},
\label{aSM}
\end{equation}
where $a_{\mathrm{N}} = GM/r^2$
denotes the standard Newtonian acceleration.

The above expression clearly shows that the deviation from Newtonian
gravity is fully controlled by the non-extensive parameters $R$
and $\delta$. Assuming $\delta>0$, the deformation parameters determine whether the generalized entropy strengthens or weakens the gravitational interaction. Specifically, values of $R/\delta$ larger than unity increase the effective acceleration compared to the Newtonian case, while values below unity produce the opposite effect. The standard Newtonian limit is recovered when the two parameters coincide, namely for $R=\delta$. Moreover, the quantity controlling the deviation from Newtonian gravity grows with the radial distance. As a result, the effects of the generalized entropy become increasingly relevant at large scales, where departures from the standard gravitational behavior may be amplified. This behavior is consistent with the broader picture emerging from generalized entropic approaches to gravity, according to which non-extensive effects may predominantly manifest themselves in the IR regime of gravitational interactions~\cite{Sheykhi:2010wm,verlinde2011origin,Tsallis2013,Czinner2016,Moradpour2018}.

The corresponding gravitational potential is obtained from the standard relation
$\vec{F}=-\nabla\Phi$. For a spherically symmetric configuration, this relation reduces to
\begin{equation}
\Phi_{\mathrm{SM}}(r)=
-\frac{G Mm}{2 r R}
\left[
(2R - \delta)\,
{}_2F_1\!\left(
-\frac{1}{2}, -\frac{R}{\delta}; \frac{1}{2};
-\frac{\pi r^2 \delta}{\ell_p^2}
\right)
+ \delta \left(1 + \frac{\pi r^2 \delta}{\ell_p^2}\right)^{R/\delta}
\right]. \label{Gravitypot}
\end{equation}
where ${}_2F_1(a,b;c;z)$ is the Gauss hypergeometric function.

In the case $R=\delta$, the second parameter of the Gauss hypergeometric function becomes $-1$, so that it truncates to the first-order polynomial \begin{equation} 
{}_2F_1\!\left( -\frac{1}{2},-1;\frac{1}{2}; -\frac{\pi r^2\delta}{\ell_p^2} \right) = 1-\frac{\pi r^2\delta}{\ell_p^2}. \end{equation} 
Substituting this result into Eq.~(\ref{Gravitypot}), one recovers the standard Newtonian gravitational potential, $\Phi_{\mathrm{N}}(r) = -\frac{GMm}{r}$.

On the other hand, in the limit $R\rightarrow0$, one can show that Eq.~(\ref{Gravitypot}) consistently reduces to the gravitational potential associated with the R\'enyi entropy framework,
\begin{equation}
\Phi_{\mathrm{R}}(r)
=
-GMm
\left[
\frac{1}{r}
+
\frac{\sqrt{\pi\delta}}{\ell_p}
\tan^{-1}\!\left(
\frac{\sqrt{\pi\delta}\,r}{\ell_p}
\right)
\right].
\end{equation}
%For negative values of $\delta$, one gets
%\begin{gather}
%    \Phi_{\text{R\'{e}nyi}}(r)=G Mm \left(
%    \frac{1}{r}
%    -\frac{\sqrt{-\pi \, \delta}\,\tanh\!\left(\frac{\sqrt{\pi \, \delta}\, r}{\ell_p}\right)}{\ell_p}
%    \right).
%\end{gather}

To further characterize the gravitational effects induced by the SM correction, it is useful to associate the modified potential with an effective mass density through the Poisson equation. Since the latter involves the gravitational potential per unit test mass, one has $\rho(r)=
\frac{1}{4\pi G}\nabla^2\Phi(r)$.
For a spherically symmetric configuration, using Eq.~(\ref{Gravitypot}) one obtains, for $r>0$,
\begin{equation}
\rho_{\mathrm{SM}}(r)
=
\frac{
M\left(\frac{R}{\delta}-1\right)\delta
\left(
1+\frac{\pi r^{2}\delta}{\ell_p^{2}}
\right)^{R/\delta-2}
}{
2r\ell_p^{2}
}\,.
\label{rhoSM}
\end{equation}
This result admits a useful physical interpretation. In the standard Newtonian case, the region outside a pointlike gravitating source is vacuum, so that the corresponding density vanishes for $r>0$. By contrast, the SM correction gives rise to a non-vanishing effective density outside the source. This density should be understood as an effective contribution induced by the modified entropic structure, rather than as an ordinary matter distribution.

In this sense, the generalized entropy correction can mimic a distributed gravitational component around the central mass. Its radial profile is controlled by the deformation parameters $R$ and $\delta$ and by the area-dependent factor entering the entropy correction. This provides a natural way to interpret the modification of the force law in terms of an effective source in the Poisson equation.

Furthermore, the sign of the effective density is determined by the ratio $R/\delta$. For $\delta>0$, the Newtonian limit $R=\delta$ gives $\rho_{\mathrm{SM}}=0$ for $r>0$, consistently recovering the standard vacuum solution outside the source. For $R>\delta$, the effective density is positive and corresponds to an additional attractive contribution. Conversely, for $R<\delta$, the effective density becomes negative, signaling a weakening of the gravitational attraction relative to the Newtonian case and indicating that such parameter regions may be less favorable for mimicking an additional matter component.
Thus, the non-vanishing density outside the source may be viewed as an effective description of the way in which the non-extensive entropy modifies the gravitational response of the system. Rather than introducing new matter degrees of freedom, the SM framework encodes these corrections through the deformation parameters $(R,\delta)$.

\subsection{Emergence of MOND-like Behavior from SM Entropy}
The large-distance behavior of the SM acceleration \eqref{aSM} provides a natural framework to investigate possible connections with IR modifications of gravity. In particular, it is instructive to examine whether the asymptotic scaling predicted by the SM entropy can reproduce the deep-MOND regime. To this end, we analyze the large-distance limit of Eq.~(\ref{aSM}) and compare the resulting behavior with the standard MOND acceleration law. This comparison not only identifies the conditions under which MOND-like dynamics emerge, but also allows one to relate the MOND acceleration scale $a_0$ to the SM deformation parameters and the gravitating mass.

At large distances satisfying $\delta\frac{\pi r^{2}}{\ell_p^{2}}\gg1$, Eq.~(\ref{aSM}) behaves as
$a_{\mathrm{SM}} \propto r^{2R/\delta-4}$.
In order to reproduce the deep-MOND behavior, $a(r)\propto 1/r$ \cite{milgrom1983mond,milgrom1983laws}, we then impose $2\frac{R}{\delta}-4=-1$,  
which finally gives 
\begin{equation} \frac{R}{\delta}=\frac{3}{2}. 
\label{mondcondition} 
\end{equation} 
Thus, the SM entropy naturally leads to a MOND-like behavior when the deformation parameters satisfy Eq.~(\ref{mondcondition}).

MOND has proven remarkably successful in accounting for several empirical regularities observed in galaxies without invoking particle dark matter. In particular, it naturally explains the asymptotic flattening of galaxy rotation curves, the baryonic Tully--Fisher relation, and the tight correlation between the observed acceleration and the baryonic mass distribution~\cite{milgrom1983mond,milgrom1983laws,mccaugh2011btfr,benkrane2025jeans}. These achievements have established MOND as an effective phenomenological description of gravity in the weak-acceleration regime~\cite{famaey2012mond,bekenstein2004teves}. In this context, the fact that the SM entropy reproduces a MOND-like behavior for a specific choice of the deformation parameters suggests that the observed IR modifications of galactic dynamics may admit a thermodynamic interpretation. From this perspective, the SM framework may encode, at an effective level, microscopic properties of the gravitational degrees of freedom that manifest themselves through non-extensive corrections to the entropy--area relation.

The condition (\ref{mondcondition}) further allows us to relate the SM deformation parameters to the MOND critical acceleration scale $a_0$. Substituting Eq.~(\ref{mondcondition}) into the asymptotic expression of the SM acceleration, one finds
\begin{equation} 
a_{\mathrm{SM}} \simeq \frac{GM}{r \ell_p} \sqrt{\pi\delta}. 
\end{equation}
Comparing this result with the deep-MOND acceleration law $a(r)=\frac{\sqrt{GMa_0}}{r}$ \cite{milgrom1983laws,milgrom1983mond},
one obtains 
\begin{equation} 
a_0= \frac{GM\pi\delta}{\ell_p^2} = \frac{2GM\pi R}{3\ell_p^2}. \label{critical} 
\end{equation}
This relation shows that the MOND acceleration scale can be effectively reproduced within the SM framework, provided that the deformation parameters are tied to the gravitating mass. In particular, assuming the observational value
$a_0\simeq1.2\times10^{-10}\,\mathrm{m\,s^{-2}}$, one finds
\begin{equation}
\delta
\simeq
\frac{1.496\times10^{-70}\,\mathrm{kg}}
{M(\mathrm{kg})}.
\label{finres}
\end{equation}
Thus, if $a_0$ is regarded as universal, the SM parameters should be interpreted as effective, system-dependent quantities rather than universal constants.

Interestingly, one finds that, for representative mass scales, Eq.~(\ref{finres}) yields small values of the deformation parameter. For instance, for a solar-mass object, $M=M_\odot\simeq1.99\times10^{30},\mathrm{kg}$, one obtains $\delta_\odot\simeq7.5\times10^{-101}$, while for a typical galaxy with $M\simeq10^{11}M_\odot$, this decreases to $\delta_{\mathrm{gal}}\simeq7.5\times10^{-112}$.

It is worth emphasizing, however, that the quantity governing the modification of the gravitational dynamics is not the deformation parameter itself, but rather the combination $\delta\pi r^2/\ell_p^2$, which is amplified by the enormous hierarchy between astrophysical and Planckian length scales. For instance, at a characteristic galactic distance of $r\sim10\,\mathrm{kpc}$, one finds $\delta\pi r^2/\ell_p^2\sim10^{-1}$, indicating that entropy-induced corrections may become relevant on galactic scales while remaining effectively negligible at shorter distances. From this perspective, the inferred values of the deformation parameter are fully consistent with the IR character of the MOND-like behavior emerging from the SM framework.

Interestingly, MOND-like phenomenology has also emerged in other extensions of gravitational physics, including scenarios involving quantum-gravity-induced modifications of the uncertainty principle~\cite{Gine:2020izd,Sevinc:2026lcv}. This may hint at a deeper connection between IR modifications of gravity and the microscopic structure of spacetime.

\section{Closing remarks}
\label{Conc}
In this work, we have investigated several thermodynamic and gravitational implications of the Sharma--Mittal (SM) entropy within the context of black hole physics and emergent gravity. We first examined the compatibility of the gravitational realization of the SM entropy with the Bekenstein bound, showing that the resulting consistency condition imposes nontrivial constraints on the deformation parameters. The corresponding limiting cases were found to consistently recover the previously known R'enyi and Bekenstein--Hawking results.

We then explored the consequences of incorporating the Landauer principle into black hole thermodynamics. By relating the erasure of one bit of information to the entropy variation of a Schwarzschild black hole, we derived a modified mass-loss relation whose behavior depends sensitively on the ratio $R/\delta$. The analysis of the small- and large-mass regimes revealed qualitatively distinct evaporation patterns, highlighting how generalized entropy may alter the thermodynamic response of black holes to information processing.

Within Verlinde's entropic gravity framework, we obtained the modified force, acceleration, and gravitational potential associated with the SM entropy. We showed that the corresponding corrections can be interpreted in terms of an effective density distribution surrounding the gravitating source, whose sign and magnitude are determined by the deformation parameters. Moreover, the large-distance behavior of the generalized acceleration naturally reproduces a MOND-like scaling when the condition $R/\delta=3/2$ is satisfied. In this case, the MOND acceleration scale can be expressed in terms of the SM parameters and the gravitating mass, providing a possible thermodynamic interpretation of infrared modifications of gravitational dynamics.

Overall, our results suggest that the SM entropy offers a unified framework in which information-theoretic principles, black hole thermodynamics, and emergent gravitational phenomena can be investigated from a common perspective. Further studies aimed at constraining the deformation parameters through astrophysical observations, as well as deriving them from more fundamental approaches to quantum gravity, may help clarify whether generalized entropy plays an effective or fundamental role in the description of gravitational systems. These issues remain the subject of ongoing investigation and will be explored in greater detail in future work.

\acknowledgments 
The research of GGL is supported by the postdoctoral fellowship program of the University of Lleida. GGL gratefully acknowledges the contribution of the LISA Cosmology Working Group (CosWG), as well as support from the COST Actions CA21136 - \textit{Addressing observational tensions in cosmology with systematics and fundamental physics (CosmoVerse)} - CA23130, \textit{Bridging high and low energies in search of quantum gravity (BridgeQG)} and CA21106 -  \textit{COSMIC WISPers in the Dark Universe: Theory, astrophysics and experiments (CosmicWISPers)}.

\bibliographystyle{apsrev4-1}
\bibliography{BLACKHOLE}
\end{document}